\documentclass{PoS}
\title{GCR intensity during the sunspot maximum phase\\
and the inversion of the heliospheric magnetic field}

\ShortTitle{GCR intensity during sunspot maximum phase and HMF inversion}

\author{\speaker{M. Krainev}, G. Bazilevskaya, M. Kalinin, A. Svirzhevskaya, N. Svirzhevsky\\
Lebedev Physical Institute, Moscow, Russia\\
E-mail: \email{mkrainev46@mail.ru},\\
\email{gbaz@rambler.ru}, \email{mkalinin@fian.fiandns.mipt.ru},\\
\email{svirzhak@fian.fiandns.mipt.ru}, \email{svirzhev@fian.fiandns.mipt.ru}
}

\abstract{
The maximum phase of the solar cycle is characterized by several interesting features in the solar activity, heliospheric characteristics and the galactic cosmic ray (GCR) intensity.
Recently the maximum phase of the current solar cycle (SC) 24, in many relations anomalous when compared with solar cycles of the second half of the 20-th century, came to the end. The corresponding phase in the GCR intensity cycle is also in progress.

In this paper we study different aspects of the sunspot, heliospheric and GCR behavior around this phase. First, the amplitudes of the SC 24 in the solar activity and GCR intensity are considered and compared with the previous sunspot cycles. Second, the inversion of the heliospheric magnetic field is studied for SC 21--24 using the suggested classification of its polarity distributions. Third, the GCR-effects specific for the maximum phase of SC 21--24 are considered and correlated with the behavior of the heliospheric magnetic field's strength and with the inversion of its polarity.

{\underline{Our main conclusions}} are as follows:
\begin{enumerate}
\item
The maximum phase of the sunspot SC 24 ended in 06.2014, the development of the sunspot cycle being similar to those of SC 14, 15 (the Glaisberg minimum). The maximum phase of SC 24 in the GCR intensity is still in progress.
\item
The inversion of the heliospheric magnetic field consists of three stages, characterized by the appearance of the global heliospheric current sheet (HCS), connecting all longitudes. In two ``transition dipole'' stages beside the global HCS there are additional local HCSs, while the ``inversion'' stage lies between two ``transition dipole'' ones and there is no global HCS in this stage. The ``inversion'' stage of the current SC 24 is the longest when compared with those for SC 21-23. The second ``transition dipole'' stage and hence the whole inversion period of the heliospheric magnetic field in SC 24 provisionally ended in 08.2014.
\item
The behavior of the GCR intensity in the period of the sunspot maximum phase and the inversion of the heliospheric magnetic fields for SC 21-23 demonstrates all the characteristic features for this period: the two-gap structure corresponding to two-peak structure in the sunspot activity, and the energy hysteresis. In the current SC 24 the GCR intensity shows rather unusual features and we should wait for one or even two years to see the whole picture.
\end{enumerate}
}
\FullConference{The 34th International Cosmic Ray Conference,\\
		30 July- 6 August, 2015\\
		The Hague, The Netherlands}
\begin{document}

\section{Introduction}
\noindent The maximum phase of the solar cycle is characterized by several interesting features in the solar activity, heliospheric parameters and the GCR intensity. First, the sunspot area and the heliospheric magnetic field (HMF) strength are at their highest levels during these periods and both often demonstrate the two--peak structure with the Gnevyshev Gap (GG) between the peaks (see \cite{Storini_Felici_NuovoCmento_17C_697_1994,Bazilevskaya_etal_SP_197_157-174_2000,Bazilevskaya_etal_SSR_186_359_2014} and references therein). In this paper by a maximum phase in the sunspot activity cycle we mean the period between two peaks of the two-peak structure in the Carrington rotation averaged sunspot area smoothed with 1 year period. Second, the inversion of the high-latitude solar magnetic fields (SMF) occurs in this phase and it changes the distribution of the HMF polarity in the heliosphere. Earlier we suggested the classification of the HMF polarity distributions to clarify the meaning of the HMF inversion \cite{Krainev_Kalinin_33ICRC_317_2013}. Third, as the GCR intensity in general anticorrelates with the sunspot area and HMF strength, this intensity is rather low in these periods and it demonstrates the two-gap structure corresponding to the two-peak structure of the sunspot area and HMF strength with somewhat different behavior for the low and high energy particles (the energy hysteresis, see the references in \cite{Moraal_ICRC14_11_3896-3906_1975,Krainev_Bazilevskaya_ASR_35_2124-2128_2005,Krainev_Kalinin_33ICRC_317_2013}). Our cosmic-ray group in the Lebedev Physical Institute has been studying the complex of these phenomena for more than 40 years, traditionally connecting the GCR intensity behavior with the SMF inversion (see the references in \cite{Krainev_Kalinin_33ICRC_317_2013}).

The maximum phase of the current SC 24, in many relations anomalous when compared with solar cycles of the second half of the 20-th century, has recently ended. The corresponding phase in the GCR intensity cycle is also in progress.The GCR behavior can be very specific in the maximum phase of this unusual solar cycle.

In this paper we first discuss the development of the current SC 24 and compare the observed maximum  sunspot area and minimum GCR intensity with those expected. Then we remind our classification of the HMF polarity structures aimed to clarify the meaning of the HMF inversion \cite{Krainev_Kalinin_33ICRC_317_2013} and discuss the HMF inversions in SC 21-24. Finally, we reconsider our old scenario of the long-term behavior of the solar and heliospheric parameters and the GCR intensity and correlation between them in the periods of high solar activity and the inversions of the large-scale HMF, also looking for the peculiarities in the current solar cycle.

As a proxy for the sunspot activity we use the Greenwich-USAF sunspot area $S_{ss}(t)$ from \cite{Sss_Site} and for period before 1870-s the reconstructed data from \cite{Sss_Site_Extended}. The SMF characteristics (the quasi-tilt $\alpha_{qt}$ of the heliospheric current sheet (HCS), the high-latitude line-of-sight SMF $B^{pol}_{ls}$ and the sets of the spherical harmonic coefficients $\left\{g_{lm},h_{lm}\right\}$) are from the site of the Wilcox Solar Observatory (WSO; Stanford University, USA) \cite{WSO_Site}. The HMF strength near the Earth $B^{hmf}$ is from \cite{OMNI_Site}. As  proxies for the intermediate and low energy GCR intensities, $J^{int}$ and $J^{low}$, we use the results of our stratospheric regular balloon monitoring (RBM) of cosmic rays \cite{Bazilevskaya_Svirzhevskaya_SSR_85_431-521_1998,Stozhkov_etal_Preprint_LPI_14_2007}:  the count rates of the omnidirectional counter in the Pfotzer maximum in Murmansk $N_{RBM}^{Mu}$ as a proxy for $J^{int}$ (the effective  energy $T_{eff}^{int}\approx 3$ GeV) and the difference between $N_{RBM}^{Mu}$ and the same characteristic in Moscow $N_{RBM}^{Mo}$ as a proxy for $J^{low}$ ($T_{eff}^{low}\approx$  few hundreds of MeV). As a proxy for the high energy GCR intensity, $J^{high}$, the neutron monitor data (Moscow, the effective energy $T_{eff}^{high}\approx 15$ GeV, \cite{NM_Moscow}) are used.

\section{The maximum phase of SC 24 vs previous solar cycles}
\begin{figure}[h]
\begin{center}
\includegraphics[width=1.\textwidth]{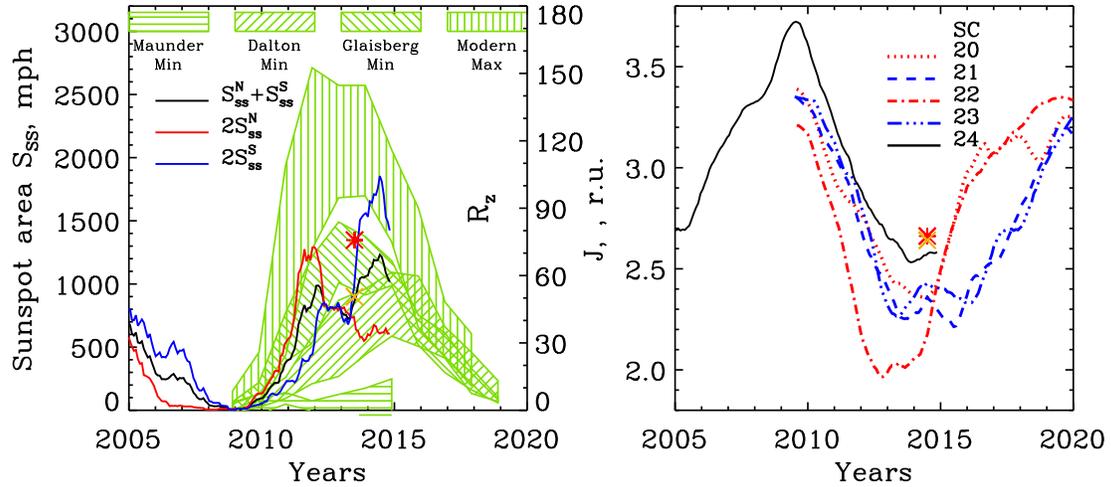}
\end{center}
\caption{\label{Sss_Nmu_24vsPrev}
{SC 24 in comparison with the previous solar cycles. All the data are smoothed with 1-year period. \underline {Left panel}}: the black curve is for the total sunspot area, while the red and blue lines are for the doubled sunspot area in the N-- and S-- hemispheres, respectively. The bands of different green shading show the development of the sunspot area averaged over the groups of the global minima and maxima of solar activity in the past and shifted to the time of SC 24. {\underline {Right panel}}: the solid black curve is for the 1-year smoothed GCR intensity $J^{int}$ and the lines of different styles show the time behavior of the same intensity for SC 20-23, shifted to the time of SC 24. The asterisks (red) and crosses (yellow) show our estimates of the characteristics in the maximum of SC 24 for two variants of the inflection point method (see the text).
}
\end{figure}

\noindent In Fig. \ref{Sss_Nmu_24vsPrev} we compare the sunspot area and the GCR intensity time profiles, $S_{ss}(t)$ and $J(t)=J^{int}(t)$, in SC 24 with their behavior in the previous solar cycles: for $S_{ss}$ with the average sunspot area for the groups of cycles of high and low activity (see \cite{Schove_1983}) and for GCR with SC 20--23. The beginnings of the previous cycles are shifted to the start of SC 24. Beside the total sunspot area the doubled sunspot areas in the N-- and S--hemispheres are shown which demonstrates what would be the total sunspot area if only N-- or S--hemisphere worked. It can be seen that the maximum phase in the sunspot area started in 02.2012 and came to the end in 06.2014, the first of two peaks being due to the sunspot activity in the N-hemisphere, while the second peak was due entirely to the activity in the S-hemisphere. As to the GCR intensity, its maximum phase is still in progress, only the first of two gaps being formed in 11.2013 (see also Fig. \ref{SMF_HMF_GCR_in_1965-2015}). The development of SC 24 demonstrates that its sunspot activity is similar to that of the cycles belonging to the Glaisberg minimum (SC 14, 15; $\approx$ 1900-1920), while the minimum GCR intensity is significantly higher than in the previous 4 cycles.

In \cite{Krainev_LPIBC_40_1_1_2013,Krainev_Kalinin_33ICRC_317_2013} the correlation was studied between the maximum/minimum values of the sunspot area and the GCR intensity and their values in the first inflection points (where $\partial P/\partial t$ is at extremum and $\partial ^2P/\partial t^2=0$ for any cyclic characteristic $P$ and time $t$). In \cite{Krainev_Kalinin_33ICRC_317_2013} the regressions between the maximum values of $S_{ss}$, $S_{ss}^{max}$, and its values in the first inflection point, $S_{ss}^{inf}$, and between $J^{min}$ and its values in the inflection points, $J^{inf}$, were found for the previous solar cycles with rather high correlation coefficients ($\rho_{ss} = 0.83$ and $\rho_J = 0.94$, respectively). Using this regressions we can find estimates for the extremum sunspot area and GCR intensity in SC 24, $S_{ss}^{24,max}$ and $J^{24,min}$. In Fig. \ref{Sss_Nmu_24vsPrev} our estimates are shown by the red asterisks (for the inflection points found from the extremum of the first time derivatives of the characteristics) and by the yellow crosses (for the inflection points found from zeros of the second time derivatives). It can be seen that both for the sunspot activity and GCR intensity our estimates poorly reproduce the observed extremum values for both variants of the inflection point method. The causes and details of this situation will be discussed elsewhere.

\section{The HMF inversion in SC 21-24}
In discussing the GCR behavior during the inversion of the HMF polarity, one usually keeps in mind the reversal of the line-of-sight component $B^{pol}_{ls}$ of the high-latitude large-scale SMF which could be followed using \cite{WSO_Site}. However, the SMF does not directly influence the GCR intensity and to understand and model the GCR intensity during such a period one should have some model on what is going on with the HMF and its polarity distribution.

One can study this distribution mainly using the WSO model which can estimate $B_r^{ss}$ on the source surface $r_{ss}=(2.5\div 3.25)r_{ph}$ in the potential approximation for two variants of the inner (at the photosphere, $r=r_{ph}$) boundary conditions: fixing from observations $B_{ls}^{ph}$ (classic variant) or $B_r^{ph}$ (radial variant) SMF components, for details see \cite{Hoeksema_PhD_1984} and references therein. So calculating $B_r^{ss}$ then finding isolines $B_r^{ss}=0$ and transporting this source surface neutral lines as the heliospheric current sheets (HCSs) to the heliosphere by the solar wind, one can find the model of the HMF polarity distribution which is in a reasonable agreement with the observed crossings of the HCSs, at least for the periods of low and intermediate solar activity.
 \begin{figure}[!t]
  \centering
  \includegraphics[width=0.75\textwidth]{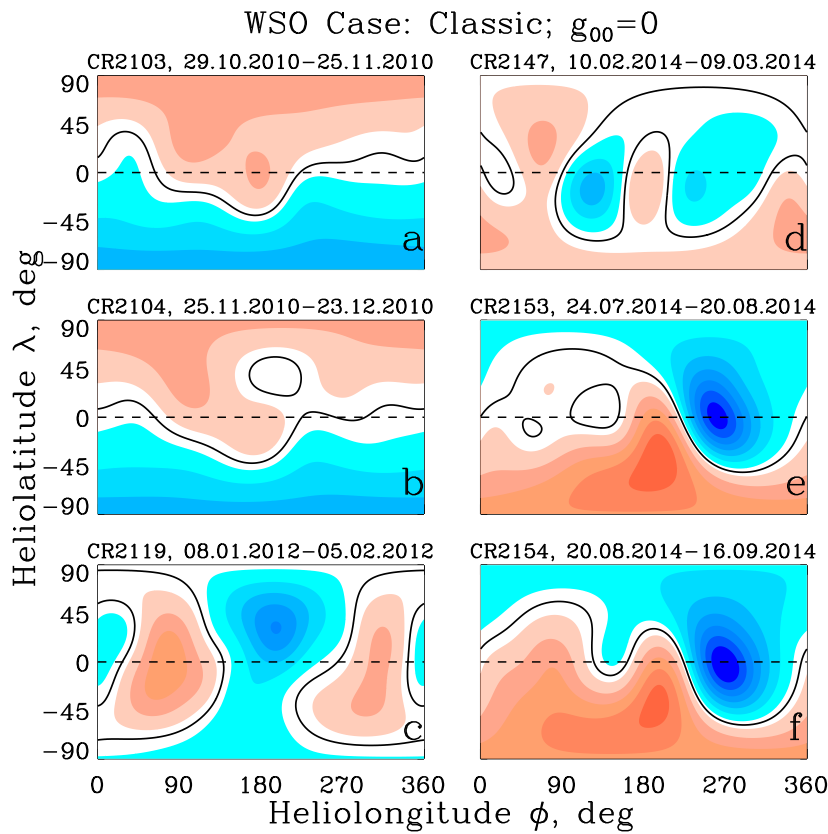}
  \caption{Three main types of the HMF polarity distribution illustrating the last HMF inversion. The thick solid lines are for the HCSs. The color (red for negative and blue for positive) stands for the HMF polarity while and its shades designate the magnitude of $B_r^{ss}$.}
  \label{CS_Maps_6_Examples}
 \end{figure}

Using the harmonic coefficients published in \cite{WSO_Site}, we calculated the above HMF polarity distribution for each Carrington rotations $N_{CR}=1642\div 2162$ (05.1976--04.2015) for two variants of the WSO model both with and without the monopole term $g_{00}$ in the sets of the spherical harmonic coefficients. Here we shall discuss only the behavior of the HMF polarity distributions for the classic variant of the WSO model without the monopole term. Note that one of the main postulates of the WSO model is the small change of the photospheric SMF during the Carrington rotation period ($\approx$ 27.28 days). This condition most easily breaks during the periods of high solar activity, so the HMF distributions for the sunspot cycle maximum phase are rather unreliable. Nevertheless, we suggest that the general picture of the HMF inversion sketched below can be at least qualitatively true (see \cite{Jokipii_Wibberentz_SSR_83_365_1998} on what HMF distribution can be expected in the solar cycle maximum phase).

\begin{figure}[h]
  \centering
  \includegraphics[width=0.75\textwidth]{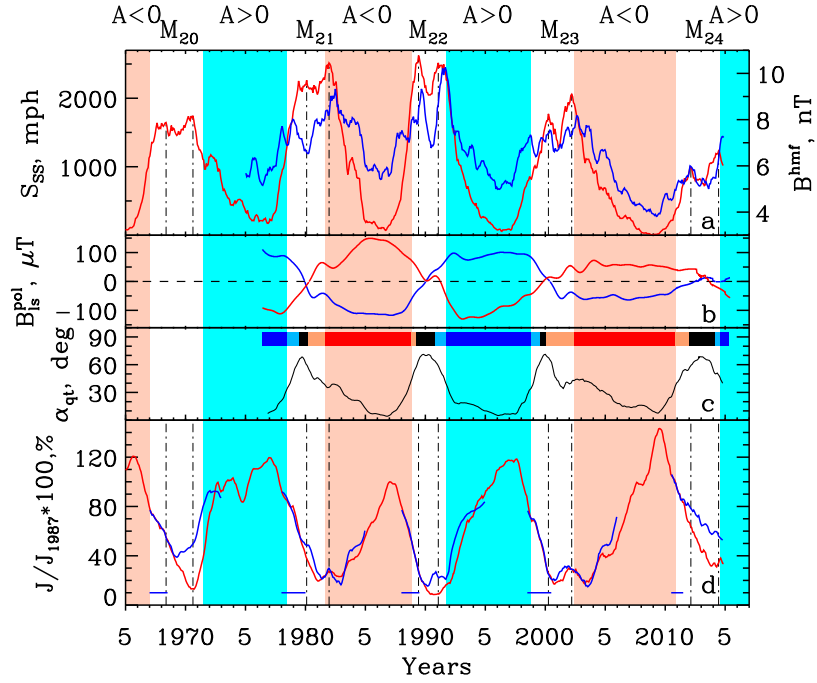}
  \caption{The solar activity, heliospheric parameters and GCR intensity in 1965-2015. All data are yearly smoothed. The periods of low solar activity with the ``dipole'' HMF polarity distributions are shaded (blue for $A>0$ and red for $A<0$) and the HMF polarity $A$ and the moments of the sunspot maxima are indicated above the panels.
  In the panels:
  (a) the total sunspot area $S_{ss}$ (red) and the HMF strength near the Earth $B^{hmf}$ (blue). The moments of two Gnevyshev peaks  for each cycle in $S_{ss}$ are shown by the vertical dash-dot lines in the panels (a) and (d); (b) the line-of-sign components of the high-latitude photospheric SMF in the N (blue) and S (red) hemispheres; (c) the HCS quasi-tilt (black) and the classification of the periods with respect to the HMF polarity states (red, black and blue sections of the horizontal band in the upper part of the panel); (d) the low energy GCR intensity normalized to 100\% in 1987 (red) and the high energy intensity (blue) for the periods of high solar activity regressed to that for the low energies by the linear regression in the periods indicated by the horizontal blue lines near the time axis.
  }
  \label{SMF_HMF_GCR_in_1965-2015}
 \end{figure}

As we showed in \cite{Krainev_Kalinin_33ICRC_317_2013} all calculated HMF polarity distributions can be divided into three types illustrated in Fig. \ref{CS_Maps_6_Examples} for the HMF inversion in the current SC 24, when the HMF dominating polarity $A$ (the sign of the HMF radial component $B_r^{hmf}$ in the N-hemisphere) changed from $A<0$ to $A>0$. The magnitude of $B_r^{ss}$ is not properly estimated in the WSO model (see \cite{Zhao_Hoeksema_SP_151_91_1994}), so correspondence between its magnitude and shades is not shown in Fig. \ref{CS_Maps_6_Examples} and we discuss only the HMF polarity and the number and forms of the HCSs. For the sake of convenience the time behavior in 1965-2015 of all related characteristics is shown in Fig. \ref{SMF_HMF_GCR_in_1965-2015}. The panels (a, f) of Fig. \ref{CS_Maps_6_Examples} illustrate the type of the HMF polarity distribution which we call the {\underline{``dipole''}} type characterized by the only and global HCS (GHCS), that is HCS connecting all longitudes both with $A<0$ (as in Fig. \ref{CS_Maps_6_Examples} (a)) and with $A>0$ (Fig. \ref{CS_Maps_6_Examples} (f)). This is the most common type of the HMF polarity distribution ($\approx 80\%$ of all time).
In panels (b, e) of Fig. \ref{CS_Maps_6_Examples} another type of the HMF polarity distribution is shown which we call the {\underline{``transition dipole''}} type and which is also characterized by the GHCS also both with $A<0$ (as in Fig. \ref{CS_Maps_6_Examples} (b)) and with $A>0$ (Fig. \ref{CS_Maps_6_Examples} (e)), but beside it one or several local HCSs exist. This type is much less common than the ``dipole'' type ($\approx 10\%$).
Finally, we call the {\underline{``inversion''}} type the third type of the HMF polarity distribution illustrated in the panels (c, d) of Fig. \ref{CS_Maps_6_Examples}. It is characterized by the absence of the global HCS with several nonglobal HCSs. This type is also much less common than the ``dipole'' type ($\approx 10\%$).

Besides this classification of the HMF polarity distribution for each Carrington rotation, we also consider the rough division of all time period into three corresponding types: 1) the ``dipole'' periods when all Carrington rotations are of the ``dipole'' type of the same $A$; 2) the ``transition dipole''  periods limited by the Carrington rotations with this type of the HMF polarity distribution also of the same $A$; and 3) the ``inversion'' periods  limited by the Carrington rotations with this type. These types of periods are marked by the differently colored sections of the horizontal bands in the upper part of Fig. \ref{SMF_HMF_GCR_in_1965-2015} (c):  the ``dipole'' periods by the dark blue (for $A>0$) or dark red ($A<0$) sections;   the ``transition dipole'' periods by the light blue ($A>0$) or light red ($A<0$) sections; and the ``inversion'' periods are marked by the black sections.

One can see in Fig. \ref{SMF_HMF_GCR_in_1965-2015} that the ``dipole'' period is characteristic for the low and intermediate solar activity and its single GHCS can be characterized by its waviness or quasi-tilt $\alpha_{qt}< \approx 40$ deg. The ``transition dipole'' period is characteristic for rather high solar activity and it corresponds to rather high quasi-tilt $\approx 40<\alpha_{qt}< \approx 60$ deg. Finally, the ``inversion'' period is characteristic for the periods of maximum solar activity and the high-latitude SMF inversions and the formally defined quasi-tilt $\alpha_{qt}> \approx 60$ deg.

As can be seen in Fig. \ref{SMF_HMF_GCR_in_1965-2015} (c) for the fast and synchronous in the N- and S-hemispheres SMF inversions (as in SC 21, 23) the HMF ``inversion'' period is also shorter, while for the prolonged and nonsynchronous SMF inversions (as in SC 22 and especially in SC 24) the HMF ``inversion'' periods are also longer. In the current SC 24 the HMF ``inversion'' period is the longest. In general the HMF ``inversion'' periods are centered with their SMF counterparts and slightly before the Gnevyshev Gap in the sunspot area and HMF strength. Usually the HMF ``inversion'' periods are surrounded by the ``transition dipole'' periods of comparable size. Note that although we provisionally marked the end of the ``transition $A>0$-dipole'' period for the current SC 24 at 08.2014 (the last Carrington rotation of this type is CR 2153), this period is too short when compared with those of the previous cycles. Due to this and because of the very low strength of the high-latitude photospheric SMF in the N-hemisphere we can expect more Carrington rotations with the ``transition $A>0$-dipole'' type of the HMF polarity distribution till the end of 2015.

\section{The GCR intensity during maximum phase and HMF inversion}
In Fig. \ref{SMF_HMF_GCR_in_1965-2015} the time profiles of all related characteristics ($S_{ss}$, $B^{hmf}$, $B^{pol}_{ls}$, the HMF polarity distribution, $\alpha_{qt}$, $J^{low}$, $J^{high}$) are shown for the last 50 years. The regular measurements of the SMF (and hence the HMF polarity distribution) started in 05.1976, while the HMF strength is shown only since 1975.

 In Fig. \ref{SMF_HMF_GCR_in_1965-2015} (a) one can easily see for SC 20-24 the double-peak structure of the sunspot maximum phase both in $S_{ss}$ and $B^{hmf}$  and corresponding double-gap signature in the GCR intensity (GG-effect). In \cite{Bazilevskaya_etal_SSR_186_359_2014} it was suggested that both step-like changes of the GCR intensity during the intermediate and low solar activity and the GG-effect around the sunspot maxima could be viewed as the manifestations of the quasi-biennial oscillations (QBO). However, note that the Gnevyshev gaps usually coincide or occur just after the inversions of the high-latitude SMF shown in Fig. \ref{SMF_HMF_GCR_in_1965-2015} (b) so there could be some physical connection between these two phenomena (see also \cite{Krainev_etal_34ICRC_439_2015} where we discuss the description of the GCR drift for all three types of the HMF polarity distribution). In Fig. \ref{SMF_HMF_GCR_in_1965-2015} (c) the quasi-tilt $\alpha_{qt}$ (classic) is shown as well as the horizontal band illustrating the suggested classification of the periods with respect to the HMF polarity (see the previous section).

Another GCR effect specific for the maximum phase of SC 20-24 and also seen in Fig. \ref{SMF_HMF_GCR_in_1965-2015} (d) is the energy hysteresis. The difference between the time profiles of the low energy intensity and high energy intensity (regressed to the low energy one for the periods before the HMF inversions) probably demonstrates the magnetic cycle \cite{Krainev_etal_ICRC19_4_481-484_1985}. When expressed as a hysteresis loop in the regression plot, the area of the loop looks greater for the even solar cycles (SC 20, 22), than for the odd ones (SC 21, 23).

In \cite{Krainev_etal_ICRC19_4_481-484_1985} we isolated two stages in the GCR intensity behavior in the maximum phase of solar cycle postulating the attenuation of $B^{hmf}$ and marking the HMF inversion by the sign of $dA/dt$ ($dA/dt>0$ for the inversion from $A<0$ to $A>0$ and vice versa). Now this scenario looks rather attractive for us if we connect the two above stages with two gaps in the double-gap structure of the GCR intensity, the postulated attenuation of $B^{hmf}$ with the first gap in this characteristic for the first stage while for the second stage we substitute the sign of $dA/dt$ for the sign of $A$ itself (because of the ``dipole'' type of the HMF polarity distribution in this period).

However, the maximum phase of the current SC 24 in the GCR intensity is still in progress. Now it looks rather unusual when compared with this phase for the previous SC 20-23. The first gap in the low energy GCR intensity occurred rather late (02.2012), while that for the high energy is still not seen at all. Nevertheless, the hysteresis loop for this even cycle looks rather wide as expected. So we should wait for about a year or even two before the situation with the maximum phase of SC 24 in the GCR intensity will be clear.

\acknowledgments
The authors thank for help the Russian Foundation for Basic Research (grants 13-02-00585, 14-02-00905)
and the Program of the Presidium of the Russian Academy of Sciences ``Fundamental Properties of Matter and Astrophysics''.

\end{document}